\documentclass[english,aip,graphicx, preprint, superscriptaddress]{revtex4}
\usepackage{lmodern}

\usepackage[T1]{fontenc}
\usepackage[latin9]{inputenc}
\usepackage{color}
\usepackage{bm}
\usepackage{graphicx}
\usepackage{esint}

\makeatletter

\usepackage{bm}

\makeatother

\usepackage{babel}

\begin{document}

\title{Effects of Line-tying on Magnetohydrodynamic Instabilities and Current
Sheet Formation}

\author{Yi-Min Huang}

\email{yimin.huang@unh.edu}

\affiliation{Space Science Center, University of New Hampshire, Durham, NH 03824}

\affiliation{Center for Integrated Computation and Analysis of Reconnection and
Turbulence}

\affiliation{Center for Magnetic Self-Organization in Laboratory and Astrophysical
Plasmas}

\author{A. Bhattacharjee}

\email{amitava.bhattacharjee@unh.edu}

\affiliation{Space Science Center, University of New Hampshire, Durham, NH 03824}

\affiliation{Center for Integrated Computation and Analysis of Reconnection and
Turbulence}

\affiliation{Center for Magnetic Self-Organization in Laboratory and Astrophysical
Plasmas}

\author{Ellen G. Zweibel}

\email{zweibel@astro.wisc.edu}

\affiliation{Department of Physics and Department of Astronomy, University of
Wisconsin, Madison, Wisconsin 53706}

\affiliation{Center for Magnetic Self-Organization in Laboratory and Astrophysical
Plasmas}
\begin{abstract}
An overview of some recent progress on magnetohydrodynamic stability
and current sheet formation in a line-tied system is given. Key results
on the linear stability of the ideal internal kink mode and resistive
tearing mode are summarized. For nonlinear problems, a counterexample
to the recent demonstration of current sheet formation by Low \emph{et
al}. {[}B. C. Low and \AA. M. Janse, Astrophys. J. \textbf{696},
821 (2009){]} is presented, and the governing equations for quasi-static
evolution of a boundary driven, line-tied magnetic field are derived.
Some open questions and possible strategies to resolve them are discussed. 
\end{abstract}
\maketitle

\section{Introduction}

In plasma physics, line-tying refers to the presence of conducting
walls enclosing the magnetized plasma of interest with a non-vanishing
component of the magnetic field normal to the wall. Line-tying is
usually employed as an idealization of the boundary condition in some
astrophysical plasmas, where the plasma density varies several orders
of magnitude over a short distance. For example, the magnetic field
lines in the coronae of stars and accretion disks are rooted in the
dense, highly conducting gas below. In the limit of infinite density
contrast, the dense gas may be treated as a conducting wall at which
the field lines are tied. 

Line-tying imposes a strong constraint on the plasma motion, therefore
in general is stabilizing. This is particularly true in the case of
ideal magnetohydrodynamics (MHD). In ideal MHD, magnetic field lines
are frozen-in to the plasma flow. If the footpoints of magnetic field
lines are not moving, i.e., the wall is rigid, then the footpoint
mapping following the field lines from one end to another is conserved.
In resistive MHD, magnetic field lines are allowed to slip though
the plasma on a resistive time scale. In this sense, the line-tying
becomes imperfect, and the stabilizing effect becomes less stringent.
It has long been speculated that solar coronal loops can remain stable
for longer than typical MHD instability time scale due to line-tying
stabilization. \cite{Raadu1972,Hood1992}

Line-tying may also be used in a broader context with a flow imposed
on the boundary, to model the convection on the solar surface or the
rotation of accretion disks. The imposed flow shuffles the footpoints
of magnetic field lines, twists up and entangles them, thereby converting
the kinetic energy of the flow into the magnetic energy in the field.
Release of the stored magnetic energy powers some of the most splendid
plasma phenomena such as solar flares and coronal mass ejections (CMEs).
Dissipation of the magnetic energy on smaller scales is thought to
be the energy source of the coronal heating. In a seminal paper, Parker
argued that when the magnetic field line topology becomes complicated
due to the footpoint shuffling, tangential discontinuities (current
sheets) inevitably will form.\cite{Parker1972} In the presence of
dissipative mechanisms such as resistivity, tangential discontinuities
will be smoothed out and the magnetic energy will be turned into heat.
Parker coined the term {}``topological dissipation'' to describe
this process. Following Parker's original suggestion, many authors
\cite{MikicSV1989,GalsgaardN1996,DmitrukGD1998,RappazzoDEV2006,RappazzoVED2007,RappazzoVED2008}
have shown by direct numerical simulation that random shuffling of
the coronal magnetic field lines by photospheric motions progressively
increases the current density, resulting in sporadic energy release.

The key ingredient of Parker's scenario of coronal heating is the
ubiquitous presence of current sheets in a magnetized plasma. To illustrate
his point, Parker considered a uniform magnetic field $B_{0}\mathbf{\hat{z}}$
line-tied to conducting plates at $z=\pm L/2$, which represent the
photosphere. Motions in the photosphere displace the field-line footpoints
and entangle the field lines. Suppose we freeze the footpoints and
let the system relax to the minimum energy state under the condition
of preserving all the ideal topological constraints. Parker argues
that in general the minimum energy state must contain tangential discontinuities.
Since its first appearance in Ref. \cite{Parker1972}, this simple
setting has become the {}``standard model'' of the Parker problem.
The property that magnetostatic equilibria tend to form tangential
discontinuities is stated as a \textquotedblleft{}Magnetostatic Theorem\textquotedblright{}
in Parker's book.\cite{Parker1994} Although Parker gives compelling
physical arguments in support of the Theorem, no rigorous proof is
given. 

Parker's claim has stimulated considerable debate that continues to
this day\cite{vanBallegooijen1985,ZweibelL1987,Antiochos1987,LongcopeS1994a,NgB1998,LongbottomRCS1998,CraigS2005},
without apparent consensus. The point of contention is whether or
not a magnetic field with neither null points, nor separatrices, nor
closed field lines can develop tangential discontinuities. It is well
established that null points, separatrices, or closed field lines
are the preferential sites for current sheet formation when a system
undergoes some instabilities or is deformed via boundary conditions.\cite{Syrovatskii1971,RosenbluthDR1973,PriestR1975,Syrovatskii1981,Low1991,CowleyLS1997,ScheperH1998}
In the Parker problem, however, none of these structures are present.
Van Ballegooijen was the first to question Parker's claim. He demonstrated
by an analytical argument that the equilibrium will always be continuous,
unless the footpoint mapping is discontinuous.\cite{vanBallegooijen1985,LongcopeC1996}
Subsequently, several authors have found smooth equilibria using either
linear theories for small footpoint displacements or Lagrangian numerical
relaxation schemes for nonlinear problems. \cite{ZweibelL1987,LongbottomRCS1998,CraigS2005,Wilmot-SmithHP2009}
However, as pointed out by Ng and Bhattacharjee,\cite{NgB1998} van
Ballegooijen's argument is valid only for simple current sheets; it
fails if a current sheet has multiple branches joined together (see
Figure 2 of Ref. \cite{NgB1998}). Ng and Bhattacharjee also proved
a theorem which asserts that there is at most one smooth equilibrium
for any given smooth footpoint mapping. It follows that if a smooth
equilibrium becomes unstable, there is no other smooth equilibrium
the system can relax to; therefore the system must relax to an equilibrium
with tangential discontinuities. The theorem of Ng and Bhattacharjee
is within the framework of reduced magnetohydrodynamics (RMHD), which
assumes the existence of a strong guide field.\cite{Kadomtsev1975,Strauss1976}
It also assumes double-periodicity in the plane perpendicular to the
guide field. It is not known at the present time whether the result
can be generalized to full MHD or more general boundary conditions,
although Aly \cite{Aly2005} appears to have taken an important step
towards that goal. Lagrangian relaxation studies indeed suggest current
sheet may form under certain circumstances,\cite{LongbottomRCS1998,CraigS2005}
although it is rather difficult to draw decisive conclusions from
numerical studies due to limits of spatial resolution.

This paper gives an overview of our recent progress
on MHD stability and current sheet formation in a line-tied system.
Previously published results are summarized in Sec.
\ref{sec:Linear} and Sec. \ref{sec:nonlinear}, with new perspectives
added. Sec. \ref{sec:Linear} summarizes the results
on linear stability regarding line-tying effects on the ideal internal
kink mode and resistive tearing mode. Parker's problem on current
sheet formation is discussed in Sec. \ref{sec:nonlinear}. The discussion
is divided into two parts. The first part discusses why the nonlinear
evolution of ideal kink mode is a relevant setting to test Parker's
hypothesis. The second part is a discussion about the recent demonstration
of current sheet formation by Low \emph{et al}.
\cite{Low2006a,Low2007,JanseL2009,LowJ2009} and our objection to
it.\cite{HuangBZ2009} Some outstanding open questions will be discussed
along the way. Sec. \ref{sec:QSevol} contains the
new results, where the governing equations for quasi-static evolution
of a boundary driven, line-tied magnetic field are derived, and a possible strategy of applying these equations
to fully settle an open question regarding the Parker problem as well
as the demonstration of Low \emph{et al}.
is outlined. Finally, some future perspectives
are discussed in Sec. \ref{sec:openissue}.

\section{Line-tying Effects on Ideal and Resistive MHD Instabilities\label{sec:Linear}}

\subsection{Ideal Internal Kink Mode}

A line-tied screw pinch is often employed as a model of solar coronal
loops. Assuming that the plasma pressure is negligible, in cylindrical
coordinates $(r,\phi,z)$, the equilibrium magnetic field is \begin{equation}
\mathbf{B}=B_{\phi}(r)\bm{\hat{\phi}}+B_{z}(r)\mathbf{\hat{z}},\label{eq:B-field}\end{equation}
which satisfies the force balance equation\begin{equation}
-\frac{d}{dr}\left(\frac{B^{2}}{8\pi}\right)-\frac{B_{\phi}^{2}}{4\pi r}=0.\label{eq:equilibrium}\end{equation}
In a periodic pinch of length $L$, we may decompose the plasma displacement
$\bm{\xi}$ into Fourier basis in both $z$ and $\phi$ directions.
Therefore we consider an eigenmode of the form $\bm{\xi}=\bm{\xi}(r)e^{\gamma t+i(kz+m\phi)}$,
where $k=2n\pi/L$, $\gamma$ is the growth rate, and both $n$ and
$m$ are integers. The $m=\pm1$ internal kink mode is unstable if
the safety factor $q\equiv2\pi rB_{z}/LB_{\phi}$ drops below unity
within the plasma and increases with radius. Unstable modes have helical
symmetry, depending on $z$ and $\phi$ in the combination $kz\pm\phi$,
and possess a so-called resonant surface $r=r_{s}$, on which $\mathbf{k}\cdot\mathbf{B}=kB_{z}\pm B_{\phi}/r_{s}=0$.
Due to the constraint of periodicity, the resonant surface must also
be a rational surface on which magnetic field lines are closed. The
driver of the instability lies within $r_{s}$, and the radial component
of the plasma displacement nearly vanishes outside it. It has been
shown that the thin transition layer near $r_{s}$ predicted from
linear theory corresponds to an infinite current sheet in finite amplitude
theory, at least within the framework of reduced, ideal MHD.\cite{RosenbluthDR1973,Waelbroeck1989}
In a line-tied pinch, the boundary condition constrains the plasma
displacement $\bm{\xi}$ to zero at $z=\pm L/2$. Therefore a Fourier
decomposition can no longer be done in $z$, and the eigenvalue problem
is inherently two-dimensional (2D). At a fundamental level, there
is a deeper issue. In a line-tied screw pinch, all magnetic field
lines are topologically equal; they all go from one end plate to another
(assuming $B_{z}\neq0$ everywhere). Therefore the notions of rational
surfaces and resonant surfaces are no longer valid. A question is:
is there still a steep gradient in the eigenmode, as in the periodic
counterpart? And if the answer is yes, where will the steep gradient
be? 

\begin{figure}
\begin{centering}
\includegraphics[scale=0.9]{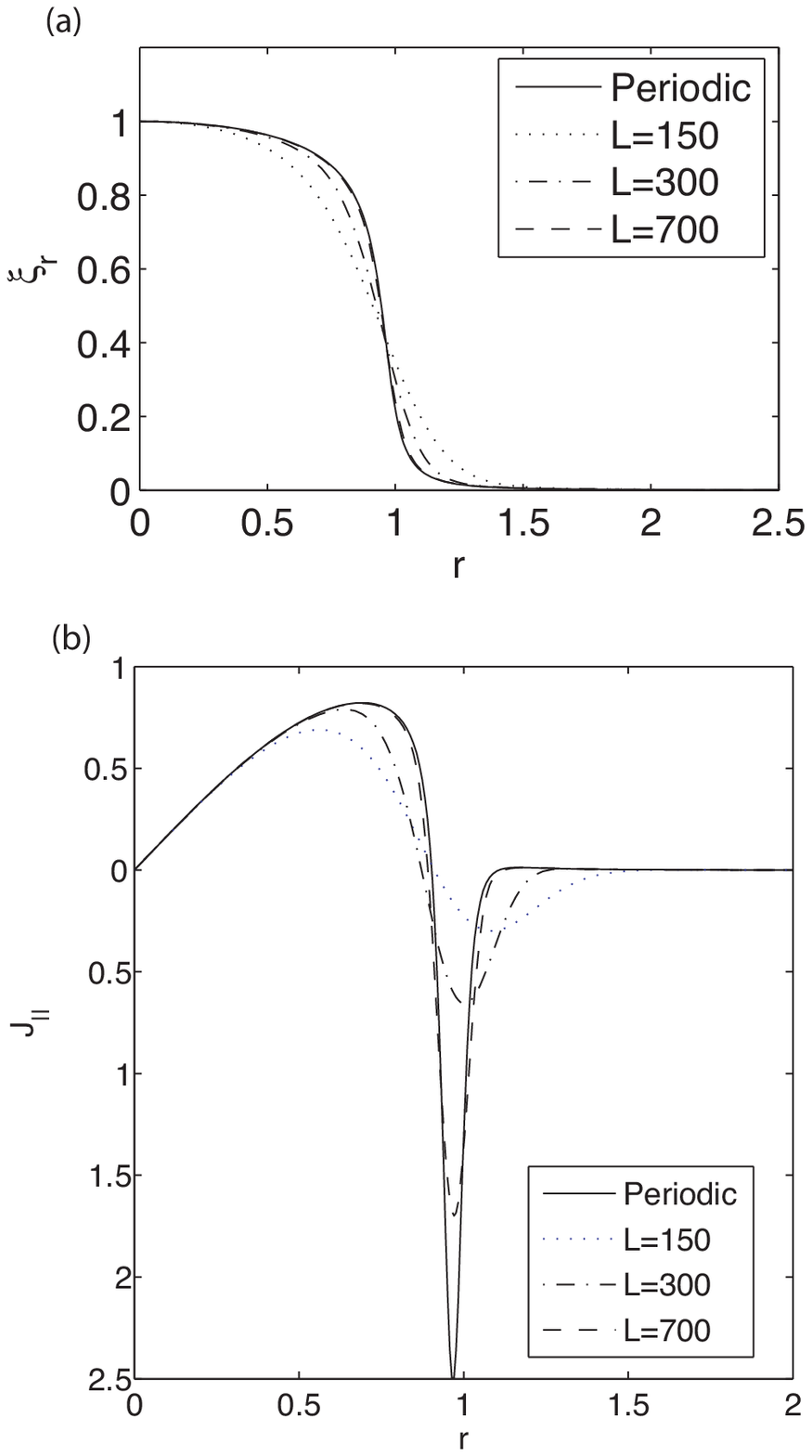}
\par\end{centering}

\caption{The radial displacement $\xi_{r}$ and parallel current $J_{\parallel}$
of eigenmodes for different $L$, along the midplane $z=0$. Also
shown for comparison is the periodic case.\label{fig:kink-eigen}}

\end{figure}

In Ref. \cite{HuangZS2006} a detailed study of the line-tied internal
kink mode has been carried out. It was found that the fastest growing
internal kink mode in a line-tied system possesses a steep internal
layer. The internal layer is located at the resonant surface of the
fastest growing periodic mode in an \emph{infinitely} long system
(i.e., $k$ is a continuum). However, the internal layer in a line-tied
system is smoother than its periodic counterpart. Only in the limit
$L\to\infty$ does the former approaches the latter. This can be seen
in Figure \ref{fig:kink-eigen}, which shows the radial displacement
$\xi_{r}$ and parallel current $J_{\parallel}$ of eigenmodes for
different $L$ along the midplane $z=0$, as compared to its periodic
counterparts. Similar results have been found in Refs. \cite{EvstatievDF2006,DelzannoEF2007}
for different equilibria and Ref. \cite{SvidzinskiML2008} including
finite pressure effects.

There is a critical length $L_{c}$ for the instability. When $L<L_{c}$
the internal kink is stabilized. The critical length $L_{c}$ is approximately
given by

\begin{equation}
L_{c}\simeq2\left|V_{Az}\right|_{r_{s}}/\gamma_{0},\label{eq:estimate}\end{equation}
where $\gamma_{0}$ and $r_{s}$ are the growth rate and the resonant
surface of the fastest growing periodic mode, respectively, and $V_{Az}$
is the $z$ component of the Alfv\'en speed. This is roughly consistent
with the heuristic stability criterion that the unstable mode is stabilized
when the Alfv\'en transit time along the loop is less than the inverse
of the growth rate in a periodic system. Once the critical length
$L_{c}$ is known, the growth rate $\gamma$ for a system of length
$L$ can be determined by solving the equation \begin{equation}
\frac{L_{c}}{L}=\frac{\gamma/\gamma_{0}}{\tanh^{-1}(\gamma/\gamma_{0})}.\label{eq:gamma-L-universality}\end{equation}
In the limit $L\to\infty$, the growth rate $\gamma\to\gamma_{0}$.
The growth rate $\gamma_{0}$ in a periodic pinch scales as \begin{equation}
\gamma_{0}\sim\epsilon^{3}\left|V_{Az}\right|/a,\label{eq:gamma0}\end{equation}
where $a$ is the characteristic radius of the pinch and $\epsilon\sim B_{\phi}/B_{z}$
is the tilt angle, which typically is a smaller parameter. The critical
length $L_{c}\sim a/\epsilon^{3}$. In solar applications, often the
coronal loop length and the axial field $B_{z}$ are given, and the
azimuthal component $B_{\phi}$ is generated by applying some footpoint
twisting. Then we can ask the following question: what is the critical
twisting angle $\phi_{c}$ for the loop to become unstable? From $L_{c}\sim a/\epsilon^{3}$
we have the critical tilt angle $\epsilon_{c}\sim(a/L)^{1/3}$ when
$L$ is given. That corresponds to a critical twisting angle\begin{equation}
\phi_{c}\sim\frac{L}{a}\epsilon_{c}\sim\left(\frac{L}{a}\right)^{2/3}.\label{eq:critical-twist}\end{equation}
The precise value of the critical tilt angle $\epsilon_{c}$ depends
on the profile. For the profile $B_{\phi}=\mbox{const}\times(r/a)\exp(-(r/a)^{2})$,
which was considered in Ref. \cite{HuangZS2006}, the critical tilt
angle at $r=a$ is approximately given by \begin{equation}
\epsilon_{c}\simeq1.1\times\left(\frac{a}{L}\right)^{1/3}.\label{eq:critical_tilt}\end{equation}
The critical tilt angle depends on the aspect ratio $(a/L)$ rather
weakly. For a coronal loop with $a/L=1/100$, the critical tilt angle
$\mbox{\ensuremath{\epsilon}}_{c}\simeq0.24$, which is approximately
$14^{\circ}$. For a shorter loop with $a/L=1/10$, the critical tilt
angle is about $29^{\circ}$; whereas for an extremely long loop with
$a/L=1/1000$, the critical tilt angle is about $6.3^{\circ}$. These
numbers are in reasonable agreement with estimates obtained from balancing
the Poynting flux from the coronal base and the observed coronal loss
rate, which give a tilt angle ranging from $10^{\circ}$ to $20^{\circ}$.\cite{Parker1988,Klimchuk2006}

\subsection{Resistive Tearing Mode}

When a small but finite resistivity $\eta$ is included in the theory,
line-tying becomes imperfect. In Ref. \cite{DelzannoEF2007a}, Delzanno
\emph{et al}. studied the resistive effects on line-tied internal
kink modes. They found that when the system length $L$ is greater
than the ideal critical length $L_{c}^{ideal}$, the mode growth rate
is similar to the ideal one. However, resistivity can make a system
with $L<L_{c}^{ideal}$ unstable. Interestingly, the critical length
of a resistive system $L_{c}^{res}$ is found to be independent of
$\eta$.

In Ref. \cite{HuangZ2009}, we studied the effects of line-tying on
the resistive tearing instability in slab geometry. The equilibrium
has a strong guide field $B_{z}$ along $z$, and $B_{y}$ changes
direction across a current layer around $x=0$. Two conducting plates
at $z=\pm L/2$ provide the line-tied boundary condition. This configuration
is ideally stable, but can be unstable to the tearing mode when $\eta>0$.
The instability is studied within the framework of the RMHD equations.\cite{KadomtsevP1974,Strauss1976}
After proper normalization, the magnetic field and the velocity can
be expressed in terms of the flux function $\psi$ and the stream
function $\phi$ as $\mathbf{B}=\mathbf{\hat{z}}+\nabla_{\perp}\psi\times\mathbf{\hat{z}}$
and $\mathbf{u}=\nabla_{\perp}\phi\times\mathbf{\hat{z}}$. The governing
equations are 

\begin{equation}
\partial_{t}\Omega+\left[\phi,\Omega\right]=\partial_{z}J+\left[\psi,J\right],\label{eq:RMHD1}\end{equation}
Here $\nabla_{\perp}\equiv\mathbf{\hat{x}}\partial_{x}+\mathbf{\hat{y}}\partial_{y}$
is the perpendicular gradient; $\Omega=-\nabla_{\perp}^{2}\phi$ ,
$J=-\nabla_{\perp}^{2}\psi$ are the vorticity and the electric current;
and $\left[\psi,\phi\right]\equiv\partial_{y}\psi\partial_{x}\phi-\partial_{x}\psi\partial_{y}\phi$
is the Poisson bracket. The standard Harris sheet $B_{y}=\mbox{const}\times\tanh x$,
is assumed, but more general profiles can be easily implemented.

\begin{figure}
\begin{centering}
\includegraphics[scale=0.4]{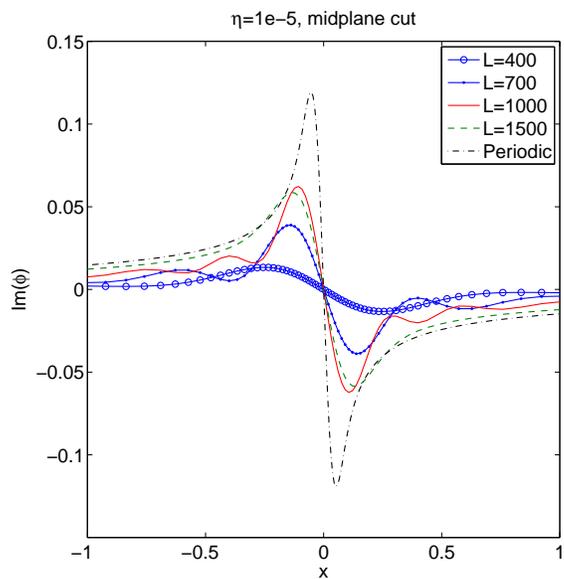}
\par\end{centering}

\caption{(Color online) The linear cut of the perturbed stream function $\tilde{\phi}$
along the mid-plane $z=0$, for line-tied cases. Also shown for comparison
is the periodic case with $k_{z}=0$.\label{fig:resisitive-layer}}

\end{figure}

In a periodic system, we may consider linear perturbations of the
form $\tilde{\phi}=\tilde{\phi}(x)e^{\gamma t+ik_{y}y+ik_{z}z}$ and
$\tilde{\psi}=\tilde{\psi}(x)e^{\gamma t+ik_{y}y+ik_{z}z}$. The system
is unstable to the tearing mode with a growth rate $\gamma\sim\eta^{3/5}$
in the asymptotic limit $\eta\to0$. With line-tying, Fourier decomposition
along $z$ is not possible and a two-dimensional eigenvalue problem
has to be solved. In this case, the system is unstable only when $L$
is greater than a critical length $L_{c}$, which is found to be independent
of $\eta$. However, near marginal stability, the linear unstable
mode grows slowly, and plasma inertia can be neglected everywhere.
With this {}``force-free'' approximation, the growth rate $\gamma$
is proportional to $\eta$. When $L$ is sufficiently long, plasma
inertia can no longer be neglected within the resistive layer. In
this regime, the line-tied growth rate approaches the periodic one,
and the $\gamma\sim\eta^{3/5}$ scaling is recovered. The transition
from the force-free regime to the inertia dominated regime occurs
at a transition length $L_{t}$ which scales as $L_{t}\sim\eta^{-2/5}$.
Notice that without line-tying, the force-free approximation only
applies in the outer region; in the resistive layer, plasma inertia
(or viscosity) has to be included, otherwise the equations become
singular at the resonant surface where $\mathbf{k}\cdot\mathbf{B}=0$.
In a line-tied system, on the other hand, it is possible that the
force-free approximation is applicable over the whole domain, provided
that the mode grows slowly. This is because that there is no resonant
surface in a line-tied system, and the equations are not singular
even without plasma inertia. The resistive layer persists even with
line-tying. However, line-tying effects make the resistive layer smoother;
when the system length $L$ becomes longer, the resistive layer becomes
steeper. This general trend, as shown in Figure \ref{fig:resisitive-layer},
resembles that in the internal layer of the ideal internal kink mode.
Similar results have been found in cylindrical geometry by Delzanno
\emph{et al}. \cite{DelzannoF2008}

\section{Current Sheet Formation and the Parker Problem\label{sec:nonlinear}}

\subsection{Nonlinear Saturation of the Ideal Kink in Line-tied Geometry}

A question that remains open is the nonlinear saturation of ideal
internal kink mode in a line-tied screw pinch. After the onset of
the instability, if we let the magnetic field relax to a minimum energy
state but preserve all the ideal topological constraints, will it
develop a tangential discontinuity? This is an excellent setting for
testing Parker's conjecture, as the initial condition contains neither
null points nor separatrices nor closed field lines. In a periodic
pinch the answer to the question is yes, as shown by Rosenbluth \emph{et
al}. that a helical ribbon-like current sheet will form at the resonant
surface, which is also a rational surface.\cite{RosenbluthDR1973,Waelbroeck1989}
In a line-tied system, the answer is not so clear. The steep gradient
in the line-tied eigenfunction (Figure \ref{fig:kink-eigen}) suggests
that a thin current filament will form. Indeed, a helical thin current
filament has been observed in several nonlinear simulations. \cite{LionelloSEV1998,LionelloVEM1998,GerrardH2003,GerrardH2004,GerrardHB2004,BrowningGHKV2008}
However, whether it is a current singularity or not is very difficult
to settle by a numerical calculation. We have seen that line-tying
in general has the effect of smoothing out the internal layer of the
linear eigenmode; it may have a similar effect to the nonlinear saturation
and turn a current singularity into a thin current layer with a finite
width. If this turns out to be the case, we may expect the current
layer to become thinner as the system length becomes longer, assuming
that the equilibrium profile remains unchanged. 

The theorem of Ng and Bhattacharjee, as discussed in the Introduction,
asserts that if a smooth equilibrium becomes unstable, the system
must relax to an equilibrium with tangential discontinuity. According
to the theorem, it appears that the ideal internal kink mode will
develop a current sheet. However, the theorem does not apply to the
present case for the following reasons. First, the theorem is proved
within the framework of RMHD, in which the ideal internal kink mode
is marginally stable. To capture the internal kink mode one has to
carry out the expansion to higher orders. This also raises the question
whether the theorem can be generalized to full MHD or not, as some
unstable modes are clearly missing in RMHD. Second, the proof assumes
double periodicity in the perpendicular plane, and a screw pinch does
not belong to this category.

\subsection{A Counterexample to the Demonstration of Current Sheet Formation
by Low\emph{ et al}. \label{sub:Counterexample}}

Recently, in a series of papers \cite{Low2006a,Low2007,JanseL2009,LowJ2009},
Low and coauthors have tried to demonstrate unambiguously the formation
of current singularities using what they call {}``topologically untwisted
fields'', which, in the context of force-free fields, is synonymous
with potential fields. That is, they limit themselves to a special
subset of force-free fields which contains no electric current. In
this case, the magnetic field can be expressed as $\mathbf{B}=\nabla\chi$
with some potential $\chi$, where $\chi$ satisfies $\nabla^{2}\chi=0$
because $\mathbf{B}$ is divergenceless. In a simply connected, compact
domain, $\chi$ is uniquely determined up to an additive constant
if the normal derivative of $\chi$ is specified on the boundary.
Therefore, a potential field is uniquely determined by prescribing
the normal component of $\mathbf{B}$ ($=\hat{\mathbf{n}}\cdot\nabla\chi$,
where $\hat{\mathbf{n}}$ is the unit normal vector) on the boundary.
In Ref. \cite{JanseL2009}, the authors consider a potential field
in a cylinder of finite length as an initial condition; the normal
component of the field is nonzero only at the top and the bottom of
the cylinder. The cylinder is then compressed to a shorter length.
Because the whole process is governed by the ideal induction equation,
the normal component of the field remains the same. By making a \emph{key}
\emph{assumption} that the field will remain potential during the
process, they circumvent the complicated problem of solving for the
various stages of quasi-static evolution, and simply calculate the
magnetic field in the final state as the potential field which satisfies
the boundary conditions. They then numerically calculate the magnetic
field line mapping from one end to the other, for both the initial
and the final states. They find that for a three-dimensional (3D)
field, the field line mapping, and hence, the topology, is changed.
Because an ideal evolution preserves the magnetic topology, the authors
claim that the field must evolve into a state which contains tangential
discontinuities; furthermore, singularities may form densely due to
the ubiquitous change of the field line mapping. 

The assumption that the field will remain potential
during the compression, however, is not proven in their work. The
reasoning behind the assumption could be understood as follows. Consider
a magnetic field line (the base field line) and a neighboring field
line which is infinitesimally separated from it. As we follow along
the field lines, in general the neighboring field line winds around
the base field line at a rate $d\phi/ds\neq0$, where $s$ is the
arc length along the field line and $\phi$ is the azimuthal angle
of the displacement vector between the two field lines relative to
a comoving reference frame. Here the comoving reference frame is specified
by three mutually perpendicular vectors at each point along the field
line. One of the vectors is tangent to the field line, while the other
two follow parallel transport along the field line. (See Appendix
\ref{sec:comovingframe} for the detail of how parallel transport
is defined. One also should not confuse the azimuthal angle $\phi$
here with the angle of rotational transform in a toroidal device,
where the base field line is a closed loop. For the rotational transform,
the well-known Frenet-Serret frame is a good reference frame. The
Frenet-Serret frame does not follow parallel transport in case the
base field line has a nonvanishing torsion. See Appendix \ref{sec:comovingframe}
for further discussion.) In general, different neighboring field lines
have different winding rates. However, for a force-free field satisfying\begin{equation}
\nabla\times\mathbf{B}=\lambda\mathbf{B}\label{eq:FFa}\end{equation}
with $\mathbf{B}\cdot\nabla\lambda=0$, it can be shown that the averaged
local winding rate over all neighboring field lines is related to
$\lambda$ as (see Appendix \ref{sec:comovingframe} for a derivation)
\begin{equation}
\left\langle \frac{d\phi}{ds}\right\rangle =\frac{\lambda}{2}.\label{eq:average_winding_rate}\end{equation}
This simple relation shows that parallel current is related to the
winding between neighboring field lines. For a potential field, which
satisfies $\lambda=0$ everywhere, the average local winding rate
$\left\langle d\phi/ds\right\rangle $ is equal to zero. This justifies
the use of the term {}``untwisted fields'' for potential fields.
Intuitively, it would appear to require a nonzero vorticity on the
boundary, which is absent in a simple expansion or compression, to
twist up the field. Therefore, it is plausible that a potential field
will remain potential during a simple expansion or compression. However,
in Ref. \cite{HuangBZ2009}, an explicit counterexample is constructed
in a two-dimensional (2D) slab geometry. The counterexample demonstrates
that a potential field can evolve to a smooth, non-potential field
under simple expansion or compression. The construction is summarized
as follows.

Consider a 2D configuration in Cartesian geometry, with $x$ the direction
of symmetry. A general magnetic field may be expressed as \begin{equation}
\mathbf{B}=B_{x}\mathbf{\hat{x}}+\mathbf{\hat{x}}\times\nabla\psi,\label{eq:2d_B}\end{equation}
where $B_{x}=B_{x}(y,z)$ and $\psi=\psi(y,z)$. For a force-free
field, $B_{x}=B_{x}(\psi)$ and $\psi$ must satisfy the Grad-Shafranov
equation\begin{equation}
\nabla^{2}\psi=-B_{x}\frac{dB_{x}}{d\psi},\label{eq:GS}\end{equation}
where $\lambda$ in Eq. (\ref{eq:FFa}) is equal to $-dB_{x}/d\psi$
in this representation. If the field is potential, then $B_{x}=\mbox{const}$
and $\nabla^{2}\psi=0$. 

Let us consider a force-free field bounded by two conducting surfaces
at $z=0$ and $z=L$, with all field lines connecting one end plate
to the other. The footpoint mapping from one end to the other is characterized
by $\psi$ on the boundaries, as well as the axial displacement along
$x$ when following a field line from the bottom to the top: \begin{equation}
h(\psi)=B_{x}(\psi)\left(\int_{z=0}^{z=L}\frac{1}{\left|\nabla\psi\right|}ds\right)_{\psi}.\label{eq:h}\end{equation}
Here $ds$ is the line element on the $y-z$ plane, and the subscript
$\psi$ indicates that the integration is done along a constant $\psi$
contour. 

Let the initial condition be $(B_{x0},\psi_{0})$ with a system length
$L_{0}$. Suppose the system has undergone a simple expansion or compression
such that the system length changes to some $L\neq L_{0}$. Since
ideal evolution preserves the footpoint mapping, the flux function
$\psi$ on the boundaries and the axial displacement $h(\psi)$ must
remain unchanged. To determine the final force-free equilibrium, we
must solve the two coupled equations \begin{equation}
\nabla^{2}\psi=-B_{x}\frac{dB_{x}}{d\psi},\label{eq:GS1}\end{equation}
and

\begin{equation}
B_{x}(\psi)=h(\psi)\left(\int_{z=0}^{z=L}\frac{1}{\left|\nabla\psi\right|}ds\right)_{\psi}^{-1},\label{eq:Bx2}\end{equation}
subject to the conditions $\psi(y,0)=\psi_{0}(y,0)$, $\psi(y,L)=\psi_{0}(y,L_{0})$,
where the axial displacement $h(\psi)$ is determined from the initial
condition. The set of coupled equations (\ref{eq:GS1}) and (\ref{eq:Bx2})
is called a generalized differential equation, which in general requires
numerical solutions. \cite{GradHS1975} 

In Ref. \cite{HuangBZ2009}, we consider the initial condition $\psi_{0}=y+\epsilon\sin(y)\cosh(z-1/2),$
$B_{x0}=1$, and $L_{0}=1$, where $\epsilon\ll1$ is a small parameter.
Eqns. (\ref{eq:GS1}) and (\ref{eq:Bx2}) are solved for arbitrary
$L$ both analytically with a perturbation theory and numerically.
The results from both approaches are consistent, and the final state
is found to be smooth, but contains non-vanishing current, and is
therefore non-potential.

It should be pointed out that our configuration is
not within a compact domain. Furthermore, the example we worked out
is periodic along the $y$ direction, even though that is not required
by the formalism. Topologically, the geometry is equivalent to the
space between two concentric, infinitely long cylinders, therefore
not simply connected.\cite{Low2009} Consequently, specifying the
normal component of the magnetic field on the boundary does not uniquely
determine the potential field. One can, for example, add a constant
$B_{y}$ or $B_{x}$ to the field without changing the normal component
of $\mathbf{B}$. In this regard, our configuration does not belong
to the same class of configurations considered by Low \emph{et
al}. Nevertheless, our example shows that the demonstration
by Low \emph{et al}. is not complete
if their key assumption is not proven. Another criticism of the demonstration
of Low \emph{et al.} has been
raised recently by Aly and Amari from a different point of view.\cite{AlyA2010}
To fully settle the issue in the future, a more general formalism
for the quasi-static evolution of a boundary driven magnetic field
is needed. This is the topic of the next section.

\section{Quasi-Static Evolution of a Boundary Driven Line-tied Magnetic Field\label{sec:QSevol}}

The governing equation for quasi-static evolution driven by boundary
motions was derived by van Ballegooijen (VB) \cite{vanBallegooijen1985}
within the framework of RMHD. This is under the assumption that the
characteristic time scales of the boundary motions are much longer
compared to the Alfv\'en transit times, therefore the system remains
force-free for all time to a good approximation. Here we generalize
the VB equation to full MHD without assuming any asymptotic ordering. 

We assume that the system satisfies the force-free condition (\ref{eq:FFa})
for all time. The magnetic field evolves according to the ideal induction
equation\begin{equation}
\partial_{t}\mathbf{B}=-\nabla\times\mathbf{E}=\nabla\times\left(\mathbf{u}\times\mathbf{B}\right),\label{eq:induction}\end{equation}
where $\mathbf{u}$ is the plasma flow and $\mathbf{E}$ the electric
field. Now suppose the magnetic field $\mathbf{B}$ and the boundary
velocity $\mathbf{u}^{b}$ are given at $t=t_{0}$. The objective
is to find a plasma flow $\mathbf{u}$ which is consistent with the
prescribed boundary flow $\mathbf{u}^{b}$, and which, when evolving
the magnetic field according to the ideal induction equation (\ref{eq:induction}),
carries the system to a neighboring force-free equilibrium at $t=t_{0}+dt$.
Following VB, first we take the time derivative of the force-free
condition (\ref{eq:FFa}). This yields \begin{equation}
\nabla\times\partial_{t}\mathbf{B}=\mathbf{B}\partial_{t}\lambda+\lambda\partial_{t}\mathbf{B}.\label{eq:VB1}\end{equation}
Using the induction equation (\ref{eq:induction}) and $\mathbf{B}\cdot\nabla\lambda=0$,
Eq. (\ref{eq:VB1}) may be rewritten as \begin{eqnarray}
\frac{d\lambda}{dt}\mathbf{B} & = & \left(\partial_{t}\lambda+\mathbf{u}\cdot\nabla\lambda\right)\mathbf{B}=\nabla\times\left(\lambda\mathbf{E}-\nabla\times\mathbf{E}\right)\nonumber \\
 & = & -\nabla\times\left(\lambda\mathbf{u}\times\mathbf{B}-\nabla\times\left(\mathbf{u}\times\mathbf{B}\right)\right).\label{eq:VB2}\end{eqnarray}
 Taking the divergence of Eq. (\ref{eq:VB2}) yields \begin{equation}
\mathbf{B}\cdot\nabla\frac{d\lambda}{dt}=0,\label{eq:dlambdadt}\end{equation}
which is consistent with the condition that $\mathbf{B}\cdot\nabla\lambda=0$
is satisfied for all time. Equation (\ref{eq:VB2}) is the equivalent
of the VB equation in the framework of full MHD. It has to be solved
for $\mathbf{u}$ subject to the boundary condition prescribed by
the boundary flow $\mathbf{u}^{b}$. In other words, we have to find
a flow $\mathbf{u}$ such that $\nabla\times\left(\lambda\mathbf{u}\times\mathbf{B}-\nabla\times\left(\mathbf{u}\times\mathbf{B}\right)\right)$
is proportional to $\mathbf{B}$ everywhere inside the domain. Clearly
only the component of $\mathbf{u}$ perpendicular to $\mathbf{B}$,
$\mathbf{u}_{\perp}$, can be determined from this approach. The parallel
component $u_{\parallel}$ is arbitrary, except that it has to be
consistent with the boundary condition. Equation (\ref{eq:VB2}) may
be solved for very general boundary conditions, including convection
($\mathbf{u}^{b}\cdot\mathbf{\hat{n}}=0$), expansion or compression
($\mathbf{u}^{b}\times\mathbf{\hat{n}}=0$), or a combination of both.
If the trajectory of each boundary point is prescribed, solving Eq.
(\ref{eq:VB2}) at each instance and integrating Eq. (\ref{eq:induction})
at the same time will carry the magnetic field through a series of
force-free equilibria. Notice that Eq. (\ref{eq:VB2}) is linear in
$\mathbf{u}$ at each instant of time. That is, if $(d\lambda_{1}/dt;\mathbf{u}_{1})$
is a solution for the boundary flow $\mathbf{u}_{1}^{b}$ and $(d\lambda_{2}/dt;\mathbf{u}_{2})$
a solution for $\mathbf{u}_{2}^{b}$, then $(c_{1}d\lambda_{1}/dt+c_{2}d\lambda_{2}/dt,c_{1}\mathbf{u}_{1}+c_{2}\mathbf{u}_{2})$
is a solution $\mathbf{u}^{b}=c_{1}\mathbf{u}_{1}^{b}+c_{2}\mathbf{u}_{2}^{b}$,
where $c_{1}$ and $c_{2}$ are arbitrary constants. 

The time evolution of $\lambda$ is of particular interest here. It
allows us to address the issue regarding Low's assumption. Starting
from a potential field, i.e., $\lambda=0$ everywhere, Eq. (\ref{eq:VB2})
becomes \begin{equation}
\frac{d\lambda}{dt}\mathbf{B}=\nabla\times\nabla\times\left(\mathbf{u}\times\mathbf{B}\right).\label{eq:VB3}\end{equation}
The field will remain potential if $d\lambda/dt=0$ everywhere. Therefore,
it will be sufficient to take the initial condition of Low \emph{et
al}., and solve Eq. (\ref{eq:VB3}). The boundary conditions can be
chosen as $\mathbf{u}_{top}^{b}=\mathbf{\hat{z}}$, $\mathbf{u}_{bottom}^{b}=0$,
and $\mathbf{u}_{side}^{b}$ varies continuously from the bottom to
the top and matches the velocity at the bottom and the top. If the
solution turns out to have $d\lambda/dt\neq0$, then the assumption
of Low \emph{et al}. is disproved. On the other hand, if Eq. (\ref{eq:VB3})
does not have a smooth solution, then current singularities will form
immediately. A third possibility is that a smooth solution exists,
and satisfies $d\lambda/dt=0$. In this case we have to follow the
time evolution through a series of potential fields until either one
of the previous two possibilities is realized. (The field cannot remain
a smooth potential field forever, as this contradicts the findings
of Ref. \cite{LowJ2009}.) If Low \emph{et al}. are indeed correct,
this approach may answer the questions regarding when, and where current
singularities will form. It is not difficult to see that in general
$d\lambda/dt$ will not be zero, as follows. The solution to Eq. (\ref{eq:VB3})
determines the components of $\mathbf{u}$ perpendicular to $\mathbf{B}$;
that is, there are two degrees of freedom at each point. And Eq. (\ref{eq:VB3})
demands that $\nabla\times\nabla\times\left(\mathbf{u}\times\mathbf{B}\right)$
be parallel to $\mathbf{B}$, which again corresponds to two independent
constraints at each point. Therefore we have the same number of unknowns
and equations. If the magnetic field is to remain potential, then
the condition $(\nabla\times\nabla\times\left(\mathbf{u}\times\mathbf{B}\right))_{\parallel}=0$
should be satisfied as well. This would correspond to a third independent
constraint at each point, and the system is then overdetermined. Of
course, the boundary motion considered by Low \emph{et al}. is a very
special one. Therefore we cannot preclude the possibility that in
their case $\nabla\times\nabla\times\left(\mathbf{u}\times\mathbf{B}\right)\parallel\mathbf{B}$
implies $\nabla\times\nabla\times\left(\mathbf{u}\times\mathbf{B}\right)=0$,
when the solution is smooth. If this can be proven to be the case,
then the assumption of Low \emph{et al}. is proven. 

The approach outlined here has been applied to the example described
in Sec. \ref{sub:Counterexample}. And indeed the solution of Eq.
(\ref{eq:VB3}) has $d\lambda/dt\neq0$, consistent with our findings
from the approach mentioned in Sec. \ref{sub:Counterexample} that
the field will evolve into a non-potential field. A similar approach
has been taken by Scheper and Hassam in their demonstration of current
sheet formation in 2D slab geometry, for an initial field with separatrices.\cite{ScheperH1998}
The initial condition of Low \emph{et al}. is fully 3D, therefore
solving Eq. (\ref{eq:VB3}) is more difficult, requiring numerical
treatment.

\section{Future Perspectives\label{sec:openissue}}

In this paper, we have reviewed some of the recent progress regarding
the effects of line-tying on MHD instabilities and current sheet formation
and presented some new results. For the problems of linear stability,
it is shown that line-tying in general has a stabilizing effect. Furthermore,
because of the absence of a resonant surface in a line-tied system,
the internal layer of the eigenfunction, although it persists, is
smoother than its periodic counterpart. For the more important nonlinear
problems, there remain two outstanding questions: the nonlinear saturation
of the line-tied ideal internal kink mode and the complete resolution
of the issue regarding the assumption made by Low \emph{et al.} in
their demonstration of current singularities. These open questions
pose significant challenges for the future, involving analysis as
well as numerical computation. The nonlinear saturation of the internal
kink mode may be attacked by some numerical force-free field solvers
that preserve the field line topology, such as Lagrangian relaxation
schemes.\cite{CraigS1986,LongbottomRCS1998,CraigS2005} If the final
equilibrium has a thin, but non-singular current layer, it may be
fully resolved with convergence, provided that the resolution is sufficient.
For the second question, a possible approach is outlined in Sec. \ref{sec:QSevol},
by solving the derived governing equations for a quasi-static, boundary
driven magnetic field. Another possible approach is solving the final
equilibrium of the compressed system by a force-free field solver,
mentioned above. In any case, we are facing a problem where the solution
may become singular, therefore a reliable mean to ascertain singularities
numerically may be needed. 

The formalism derived in Sec. \ref{sec:QSevol} also opens up the
possibility of generalizing the proof of Ng and Bhattacharjee, which
relies heavily on the VB equation, to full MHD. This will be a topic
of future investigation. 
\begin{acknowledgments}
This research is supported by the National Science Foundation, Grant
No. PHY-0215581 (PFC: Center for Magnetic Self-Organization in Laboratory
and Astrophysical Plasmas) and the Department of Energy, Grant No.
DE-FG02-07ER46372, under the auspice of the Center for Integrated
Computation and Analysis of Reconnection and Turbulence (CICART). 
\end{acknowledgments}
\appendix

\section{Comoving Frame of Reference, Parallel Transport, and a Derivation
of Eq. (\ref{eq:average_winding_rate})\label{sec:comovingframe}}

Let $\mathbf{X}(s)$ be a magnetic field line (the {}``base'' field
line) parameterized by the arc length $s$. A comoving reference frame
along the magnetic field line is specified by three mutually perpendicular
unit vectors $\{\mathbf{e}_{1}(s),\mathbf{e}_{2}(s),\mathbf{e}_{3}(s)\}$
at each point along the field line. Here we further require $\mathbf{e}_{3}$
to be the tangent vector. That is, $\mathbf{e}_{3}=d\mathbf{X}/ds=\hat{\mathbf{b}}=\mathbf{B}/B$.
In general, the basis vectors $\left\{ \mathbf{e}_{i}\right\} _{i=1,2,3}$
follow\begin{equation}
\frac{d\mathbf{e}_{i}}{ds}=\bm{\omega}\times\mathbf{e}_{i},\label{eq:rotation}\end{equation}
where $\bm{\omega}$ can be regarded as the angular {}``velocity'',
if we treat the arc length $s$ as the time variable. We adopt the
following definition of parallel transport: a comoving frame is said
to be parallelly transported along a field line if the angular velocity
is perpendicular to the magnetic field. That is, the angular velocity
satisfies $\bm{\omega}\cdot\mathbf{e}_{3}=0$. Under this condition,
the comoving frame is not rotating with respect to the direction of
the magnetic field, therefore is a good reference frame for measuring
the local winding rate of neighboring field lines. From $d\mathbf{e}_{3}/ds=\bm{\omega}\times\mathbf{e}_{3}$
and $\bm{\omega}\cdot\mathbf{e}_{3}=0$, the angular velocity is uniquely
determined as \begin{equation}
\bm{\omega}=\mathbf{e}_{3}\times\frac{d\mathbf{e}_{3}}{ds}.\label{eq:omega}\end{equation}
If a set of basis vectors is given at one point, then the comoving
frame following parallel transport at each point can be obtained by
integrating Eq. (\ref{eq:rotation}) with the angular velocity given
by Eq. (\ref{eq:omega}). Specifically, \begin{equation}
\frac{d\mathbf{e}_{1}}{ds}=-\left(\mathbf{e}_{1}\cdot\frac{d\mathbf{e}_{3}}{ds}\right)\mathbf{e}_{3},\label{eq:e1}\end{equation}
\begin{equation}
\frac{d\mathbf{e}_{2}}{ds}=-\left(\mathbf{e}_{2}\cdot\frac{d\mathbf{e}_{3}}{ds}\right)\mathbf{e}_{3}.\label{eq:e2}\end{equation}

Having set up the comoving frame, we can now define a local coordinate
system $(a_{1},a_{2},s)$ in a neighborhood of the field line: \begin{equation}
\mathbf{x}(a_{1},a_{2},s)=\mathbf{X}(s)+a_{1}\mathbf{e}_{1}+a_{2}\mathbf{e}_{2}.\label{eq:local_coordinate}\end{equation}
Now suppose there is a neighboring field line with a trajectory defined
by $\mathbf{X}'(s)=\mathbf{X}(s)+\delta\mathbf{x}(s)=\mathbf{X}(s)+a_{1}(s)\mathbf{e}_{1}+a_{2}(s)\mathbf{e}_{2}$,
where $a_{1}$ and $a_{2}$ are infinitesimal. The differential arc
length of the neighboring field line $\mathbf{X}'(s)$ is \begin{equation}
ds'=\sqrt{\frac{d\mathbf{X}'}{ds}\cdot\frac{d\mathbf{X}'}{ds}}ds=\left(1-\delta\mathbf{x}\cdot\frac{d\mathbf{e}_{3}}{ds}\right)ds,\label{eq:arclength_prime}\end{equation}
where Eqs. (\ref{eq:e1}) and (\ref{eq:e2}) are applied. The magnetic
field at $\mathbf{X}'(s)$ is \begin{equation}
\mathbf{B}'(s)=B(s)\mathbf{e}_{3}+\delta\mathbf{x}\cdot\nabla\mathbf{B},\label{eq:BB}\end{equation}
which has a magnitude \begin{equation}
B'(s)=B(s)+\mathbf{e}_{3}\cdot\left(\delta\mathbf{x}\cdot\nabla\mathbf{B}\right).\label{eq:Bnorm}\end{equation}
Using Eqs. (\ref{eq:arclength_prime}), (\ref{eq:BB}), and (\ref{eq:Bnorm})
in the field line equation $d\mathbf{X}'/ds'=\mathbf{B}'/B'$ yields
the governing equation for $\delta\mathbf{x}$:

\begin{equation}
\left(\frac{d}{ds}\delta\mathbf{x}\right)_{\perp}=\frac{1}{B(s)}\left(\delta\mathbf{x}\cdot\nabla\mathbf{B}\right)_{\perp},\label{eq:deviation}\end{equation}
where the subscript $\perp$ indicates a projection onto the plane
spanned by $\mathbf{e}_{1}$ and $\mathbf{e}_{2}$. Equation (\ref{eq:deviation})
can be written explicitly as \begin{equation}
\frac{d}{ds}\left[\begin{array}{c}
a_{1}\\
a_{2}\end{array}\right]=M(s)\left[\begin{array}{c}
a_{1}\\
a_{2}\end{array}\right],\label{eq:components}\end{equation}
where \begin{equation}
M(s)=\frac{1}{B(s)}\left[\begin{array}{cc}
(\nabla\mathbf{B})_{11} & (\nabla\mathbf{B})_{21}\\
(\nabla\mathbf{B})_{12} & (\nabla\mathbf{B})_{22}\end{array}\right],\label{eq:matrix}\end{equation}
and \begin{equation}
(\nabla\mathbf{B})_{ij}\equiv(\mathbf{e}_{i}\cdot\nabla\mathbf{B})\cdot\mathbf{e}_{j}.\label{eq:ddot}\end{equation}
The winding rate of the neighboring field line becomes apparent if
we express the deviation $\delta\mathbf{x}$ in terms of the local
polar coordinates $(r,\phi)$, which are related to $(a_{1},a_{2})$
by $a_{1}=r\cos\phi$ and $a_{2}=r\sin\phi$. It follows from Eq.
(\ref{eq:components}) that

\begin{equation}
\frac{1}{r}\frac{dr}{ds}=\frac{1}{B(s)}\left((\nabla\mathbf{B})_{11}\cos^{2}\phi+\frac{(\nabla\mathbf{B})_{12}+(\nabla\mathbf{B})_{21}}{2}\sin2\phi+(\nabla\mathbf{B})_{22}\sin^{2}\phi\right),\label{eq:drds}\end{equation}
and \begin{equation}
\frac{d\phi}{ds}=\frac{1}{B(s)}\left(\frac{(\nabla\mathbf{B})_{22}-(\nabla\mathbf{B})_{11}}{2}\sin2\phi+(\nabla\mathbf{B})_{12}\cos^{2}\phi-(\nabla\mathbf{B})_{21}\sin^{2}\phi\right).\label{eq:dphids}\end{equation}
Equation (\ref{eq:drds}) determines whether the two field lines are
diverging from or converging towards each other, and Eq. (\ref{eq:dphids})
gives the winding rate $d\phi/ds$ of the neighboring field line with
respect to the base field line. Clearly, $d\phi/ds$ depends on $\phi$.
Therefore, in general different neighboring field lines have different
winding rates. Averaging Eqs. (\ref{eq:drds}) and (\ref{eq:dphids})
over $\phi$ yields \begin{equation}
\left\langle \frac{1}{r}\frac{dr}{ds}\right\rangle =\frac{1}{2B(s)}\left((\nabla\mathbf{B})_{11}+(\nabla\mathbf{B})_{22}\right)=-\frac{1}{2B}\frac{dB}{ds},\label{eq:drds_ave}\end{equation}
and \begin{equation}
\left\langle \frac{d\phi}{ds}\right\rangle =\frac{1}{2B(s)}\left((\nabla\mathbf{B})_{12}-(\nabla\mathbf{B})_{21}\right)=\frac{\mathbf{e}_{3}\cdot\nabla\times\mathbf{B}}{2B}.\label{eq:dphids_ave}\end{equation}
Here we have made use of the condition $\nabla\cdot\mathbf{B}=0$
in the last step of Eq. (\ref{eq:drds_ave}). Equation (\ref{eq:drds_ave})
corresponds to the conservation of magnetic flux, $d(\sigma B)/ds=0$,
where $\sigma$ is the cross section of a thin flux tube. And Eq.
(\ref{eq:dphids_ave}) relates the average winding rate to the parallel
current. For a force-free field satisfying $\nabla\times\mathbf{B}=\lambda\mathbf{B},$
Eq. (\ref{eq:dphids_ave}) can be further simplified as 

\begin{equation}
\left\langle \frac{d\phi}{ds}\right\rangle =\frac{\lambda}{2},\label{eq:dphids_ave_ff}\end{equation}
and we have thus completed the derivation of Eq. (\ref{eq:average_winding_rate}).
Heuristically, Eq. (\ref{eq:drds_ave}) may be understood by consider
a circular cross section of a radius $r$. The conservation of magnetic
flux, $d(\sigma B)/ds=0$, implies $\sigma^{-1}d\sigma/ds=-B^{-1}dB/ds$.
Using this equation with the relation \begin{equation}
\sigma^{-1}\frac{d\sigma}{ds}=\frac{1}{\pi r^{2}}\oint2\pi r\frac{dr}{ds}d\phi=\left\langle \frac{2}{r}\frac{dr}{ds}\right\rangle ,\label{eq:dsigma}\end{equation}
Eq. (\ref{eq:drds_ave}) is obtained. Likewise, Eq. (\ref{eq:dphids_ave})
may be understood by considering a loop integral around the circular
cross section and using the Stokes' theorem, which yields \begin{equation}
\oint rB_{\phi}d\phi=2\pi r\left\langle B_{\phi}\right\rangle =\pi r^{2}\mathbf{e}_{3}\cdot\nabla\times\mathbf{B}.\label{eq:Bphi_ave}\end{equation}
Substituting Eq. (\ref{eq:Bphi_ave}) in \begin{equation}
\left\langle \frac{d\phi}{ds}\right\rangle =\frac{\left\langle B_{\phi}\right\rangle }{rB},\label{eq:dphids_ave1}\end{equation}
 Eq. (\ref{eq:dphids_ave}) is recovered. Our derivation puts Eqs.
(\ref{eq:drds_ave}) and (\ref{eq:dphids_ave}) on a rigorous footing. 

It is of great interest to consider the well-known Frenet-Serret comoving
frame in conjunction with the present formulism. The Frenet-Serret
frame is defined by the tangent vector $\bar{\mathbf{e}}_{3}=d\mathbf{X}/ds$,
the principal normal vector $\bar{\mathbf{e}}_{1}=\left|d\bar{\mathbf{e}}_{3}/ds\right|^{-1}\left(d\bar{\mathbf{e}}_{3}/ds\right)$,
and the binormal vector $\bar{\mathbf{e}}_{2}=\bar{\mathbf{e}}_{3}\times\bar{\mathbf{e}}_{1}$.
These vectors satisfy the Frenet-Serret relations $d\bar{\mathbf{e}}_{3}/ds=\kappa\bar{\mathbf{e}}_{1}$,
$d\bar{\mathbf{e}}_{1}/ds=-\kappa\bar{\mathbf{e}}_{3}+\tau\bar{\mathbf{e}}_{2}$,
and $d\bar{\mathbf{e}}_{2}/ds=-\tau\bar{\mathbf{e}}_{1}$, where $\kappa$
is the curvature and $\tau$ is the torsion of the magnetic field
line.\cite{MathewsW1970} The Frenet-Serret relations can be expressed
as \begin{equation}
\frac{d}{ds}\bar{\mathbf{e}}_{i}=\bar{\bm{\omega}}\times\bar{\mathbf{e}}_{i},\label{eq:Frenet}\end{equation}
with $\bar{\bm{\omega}}=\kappa\mathbf{\bar{e}}_{2}+\tau\bar{\mathbf{e}}_{3}$.
Because $\bar{\bm{\omega}}\cdot\bar{\mathbf{e}}_{3}\neq0$, the Frenet-Serret
frame does not follow parallel transport. The angular velocity of
the the Frenet-Serret frame, $\bar{\bm{\omega}}$, is related to the
parallel transport angular velocity, $\bm{\omega}$, by $\bar{\bm{\omega}}=\bm{\omega}+\tau\mathbf{e}_{3}$.
Therefore, the Frenet-Serret frame is rotating with respect to a parallelly
transported comoving frame with an angular velocity $\tau\mathbf{e}_{3}$.
If we instead measure the winding rate of a neighboring field line
by the angle $\bar{\phi}$ with respect to $\bar{\mathbf{e}}_{1}$
of the Frenet-Serret frame, then \begin{equation}
\frac{d\bar{\phi}}{ds}=\frac{d\phi}{ds}-\tau.\label{eq:Frenet_rate}\end{equation}

In a toroidal device with its magnetic axis being a closed base field
line, we are interested in the angle between the initial and final
$\delta\mathbf{x}$ after a full toroidal circuit. The average of
this angle over many circuits leads to the notion of rotational transform.
To measure this angle, we need a reference frame that goes back to
itself after a full circuit, a requirement that in general is not
satisfied by a parallelly transported comoving frame. The Frenet-Serret
frame, on the other hand, only depends on the local geometry of a
curve, therefore is well-suited for the purpose. As such, the rotation
angle after a full circuit can be expressed as a loop integral $\oint d\bar{\phi}=\oint d\phi-\oint\tau\, ds,$
which has contributions from two parts. The first part, $\oint d\phi$,
depends on the parallel current, as discussed above. However, the
second part, $-\oint\tau\, ds$, only depends on the geometry of the
magnetic axis. Consequently, a toroidal device can have a nonvanishing
rotational transform even with a vacuum field, provided that the magnetic
axis is nonplanar (therefore $\tau\neq0$). This geometric angle is
the basis of the figure-eight and other stellarators with nonplanar
axes. Interested readers are referred to Ref. \cite{BhattacharjeeST1992}
for further discussion.

\bibliographystyle{apsrev}

\pagebreak{}%

\end{document}